\documentclass[twocolumn]{aastex631}

\usepackage{tikz}
\usepackage{xcolor}
\usepackage{amsmath}
\usepackage{bm}
\usepackage{multirow}

\newcommand{\ixpe}{\textit{IXPE}}
\newcommand{\parens}[1]{\left(#1\right)}
\newcommand{\brackets}[1]{\left[#1\right]}
\newcommand*\rot{\rotatebox{90}}

\usetikzlibrary{arrows.meta}

\begin{document}

\title{Polarization Leakage and the \ixpe\ PSF}

\author[0000-0002-6401-778X]{Jack T. Dinsmore}
\email{jtd@stanford.edu}
\author[0000-0001-6711-3286]{Roger W. Romani}
\affiliation{Department of Physics and Kavli Institute for Particle Astrophysics and Cosmology, Stanford University, Stanford, California 94305}

\begin{abstract}

By measuring photoelectron tracks, the gas pixel detectors of the Imaging X-ray Polarimetry Explorer satellite provide estimates of the photon detection location and its electric vector position angle (EVPA). However, imperfections in reconstructing event positions blur the image and EVPA-position correlations result in artificial polarized halos around bright sources. We introduce a new model describing this ``polarization leakage'' and use it to recover the on-orbit telescope point-spread functions, useful for faint source detection and image reconstruction. These point spread functions are more accurate than previous approximations or ground-calibrated products ($\Delta \chi^2\approx 3\times 10^{4}$ and $4 \times 10^4$ respectively for a bright $10^6$-count source). We also define an algorithm for polarization leakage correction substantially more accurate than existing prescriptions ($\Delta \chi^2\approx 1\times 10^{3}$). These corrections depend on the reconstruction method, and we supply prescriptions for the mission-standard ``Moments'' methods as well as for ``Neural Net'' event reconstruction. Finally, we present a method to isolate leakage contributions to polarization observations of extended sources and show that an accurate PSF allows the extraction of sub-PSF-scale polarization patterns.

\end{abstract}


\section{Introduction}
\label{sec:intro}

Gas pixel detectors \citep[GPDs;][]{costa2001efficient} map the tracks of X-ray photoelectrons, providing estimates of position, energy and, via the photoelectron direction, electric vector position angle (EVPA). With this technology the Imaging X-ray Polarimetry Explorer (\ixpe) satellite \citep{weisskopf2016imaging} has opened up the field of imaging X-ray polarimetry, with polarization measurements of dozens of celestial sources in the classic soft ($\sim 2$--8 keV) X-ray band. However, estimates of the photon direction and polarization depend on the track reconstruction method used, and errors in this reconstruction can blur the resulting image and introduce subtle artifacts into polarization maps. In particular, position uncertainty tends to be larger along the the initial photoelectron direction, and this correlation produces artificial polarization halos around bright sources, termed ``polarization leakage'' \citep{bucciantini2023polarisation}.

\begin{figure*}
  \centering
  \includegraphics[width=\linewidth]{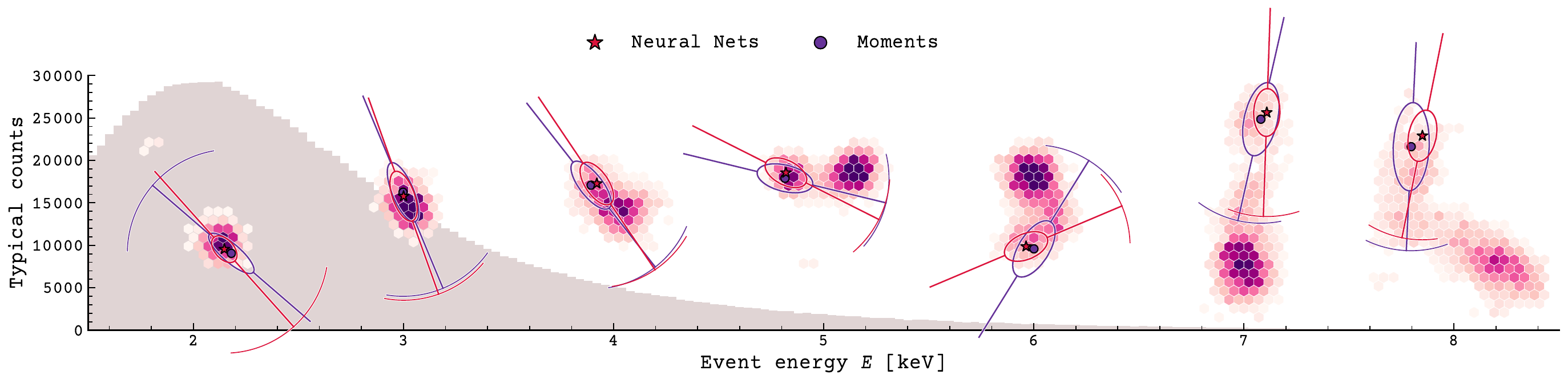}
  \caption{Real \ixpe\ level 1 tracks at different energies with reconstructed positions and EVPAs (Mom in purple, NN in red). Counts pile up in the Bragg peak at the end of the photoelectron track. The ellipses (sized $2\sigma_\parallel \times 2\sigma_\perp$) represent the characteristic uncertainty on the reconstructed position. An example spectral distribution is shown in the background to highlight that high energy tracks are valuable but rare. Error bars signifying the EVPA mean absolute error as a function of energy are shown.}
  \label{fig:tracks}
\end{figure*}

Polarization leakage amplitude depends on the reconstruction algorithm. The mission-standard moments-based (Mom) algorithm simply and robustly determines event properties from a moment decomposition of the tracks \citep{bellazzini2002novel}. An alternative algorithm uses a neural network (NN) analysis of track images, trained on simulated data, to extract event properties \citep{peirson2021deep}. NN reconstruction generally produces more accurate event weights and position estimates, with some improvement in EVPA reconstruction.

The leakage-generated Stokes parameter $Q$ and $U$ fluxes average out over the image such that no polarization is induced in large aperture measurements of point sources, but leakage does create problems for polarization mapping of extended sources. Flux distributions with sharp edges, such as the neighborhood of bright source, are particularly vulnerable to polarization leakage \citep[e.g.][]{wong2023improved, romani2023polarized}. Since X-ray polarization is often weak, leakage can have large statistical significance.

We parameterize the reconstruction errors and extract sky-calibrated telescope point-spread functions (PSFs) from \ixpe\ observations of point sources in section \ref{sec:methods}. Each of the three detector units (DU) is fed by an independent mirror module assembly (MMA) and requires a unique PSF. Using a new formalism to compute and correct for the polarization leakage induced in \ixpe\ data, we show leakage predictions that are substantially more accurate than existing methods (section \ref{sec:results}). We conclude by noting limitations of the present treatment, and a discussion of how its application can be useful for a variety of \ixpe\ measurements (section \ref{sec:conclusion}).

\section{Methods}
\label{sec:methods}

We generalize the formalism of \cite{bucciantini2023polarisation} for point source polarization leakage to apply to asymmetric PSFs in section \ref{sec:leakage}. Section \ref{sec:extended-deconvolution} gives an algorithm to extract source polarization from observations of extended sources. We describe how our sky-calibrated PSFs were extracted in section \ref{sec:psfs}, and determine leakage parameters in section \ref{sec:fit}.

\subsection{Polarization Leakage}
\label{sec:leakage}

Reconstruction error manifests as a random offset $\bm \delta$ from the true location of a photoelectron. The resulting polarization leakage patterns are controlled by the PSF of the telescope and the probability distribution of $\bm \delta$. We name the latter $P(\bm \delta|\phi)$ because it aligns with the photoelectron EVPA $\phi$ of the incident photon; tracks are elongated along the polarization axis (Fig.~\ref{fig:tracks}) which results in larger reconstruction errors.

\begin{figure}
  \centering
  \begin{tikzpicture}[scale=2.3, every text node part/.style={align=center}]
    \node[inner sep=0pt] at (0.4,0.83) {\includegraphics[width=.25\linewidth]{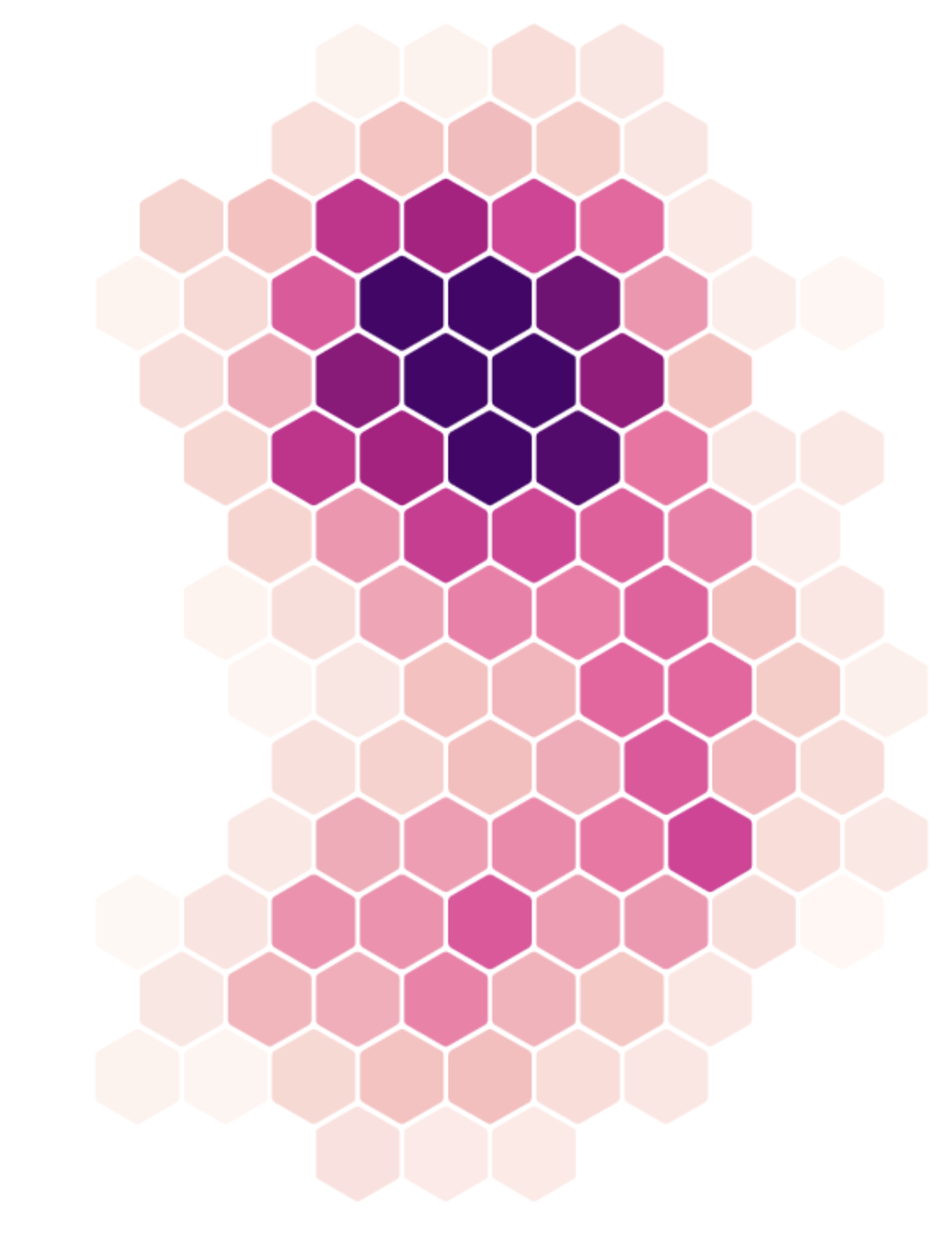}};
    \draw [very thick] plot [smooth cycle, tension=0.8] coordinates {(0, 1) (0.4, 0.4) (1, 0) (0.4, -0.4) (0, -0.7) (-0.4, -0.4) (-1, 0.1) (-0.4, 0.5)};
    \draw (0.0,0.0) node {\textbullet};
    \draw [thick] (0,0) -- node[above left] {$\bm r_0$} (0.35, 0.45);
    \fill [white, opacity=0.7] (0.52, 0.535) rectangle + (-0.137,0.15);
    \draw (0.54, 0.51) node[above left] {$\bm \delta$};
    \draw [thick] (0.35, 0.45) -- (0.6, 0.53);
    \draw (0.0,0.0) node {\textbullet};
    \draw [thick] (0,0) -- node[below right ] {$\bm r$} (0.6, 0.53);
    \draw [-{Latex[length=2.5mm]}]  (0,0) -- (0.21, 0.27);
    \draw [-{Latex[length=2.5mm]}]  (0.35, 0.45) -- (0.535,0.51);
    \draw [-{Latex[length=2.5mm]}]  (0,0) -- (0.358, 0.318);
    \draw  (-0.65,0.65) node[below] {PSF};
    \draw [-{Latex[length=2.5mm]}] (1.1,0.9) arc (30:100:0.5);
    \node at (1.1,0.7) {Bragg\\peak};
  \end{tikzpicture}
  \caption{An \ixpe\ event incident on $\bm r_0$ but observed at $\bm r$ due to reconstruction error $\bm \delta$. The observed point source is in the center.}
  \label{fig:leakage-diagram}
\end{figure}
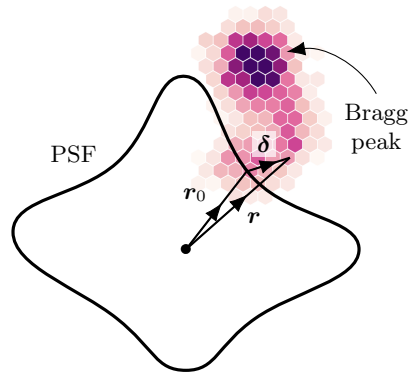

Consider a single photon from a point source incident on an \ixpe\ detector. The mirror assembly scatters the photon by the PSF, arriving at positions $\bm r_0$ with probability $P_\mathrm{mir}(\bm r_0)$. Reconstruction error causes the detected position to be $\bm r = \bm r_0 + \bm \delta$ (see Fig.~\ref{fig:leakage-diagram}). Altogether, the probability distribution for the detected position is 
\begin{equation}
  P(\bm r|\phi) = \int d^2 \bm \delta\, P(\bm \delta|\phi) P_\mathrm{mir}(\bm r - \bm \delta).
  \label{eqn:blur-psf}
\end{equation}
Once $P(\bm r|\phi)$ is computed, the average contribution of each photoelectron to the $I$, $Q$, and $U$ maps of an observation is obtained by averaging over the possible values of $\phi$:
\begin{subequations}
  \begin{align}
    I(\bm r) &= \int_0^{2\pi} d\phi\,P(\bm r|\phi)P(\phi)\\
    Q(\bm r) &= 2\int_0^{2\pi} d\phi\,P(\bm r|\phi)P(\phi)\cos(2\phi)\\
    U(\bm r) &= 2\int_0^{2\pi} d\phi\,P(\bm r|\phi)P(\phi)\sin(2\phi)
  \end{align}
  \label{eqn:iqu-def}
\end{subequations}
where 
\begin{equation}
  P(\phi) = \frac{1}{2\pi}\parens{1 + \mu_{100} \Pi \cos[2(\phi-\phi_0)]}
  \label{eqn:prob-phi}
\end{equation}
is the probability distribution of the EVPA. $\Pi$ represents the source polarization degree (PD), $\phi_0$ is the polarization angle, and $\mu_{100}$ is the detector modulation factor. The modulation factor $\mu_{100}$ describes the efficiency of a given detector and analysis scheme; it is the measured $\Pi$ for a 100\% polarized source. One divides measured $q$ and $u$ by this factor to obtain the true values. Motivated by the fact that the average of all photoelectron weights is equal to $\mu_{100}$ if the weights are optimal \citep{peirson2021toward}, we replace $\mu_{100}$ with photoelectron weights $\mu$ in this analysis.

We start by Taylor expanding the integrand of Eq. \ref{eqn:blur-psf} in $\bm \delta$ to fourth order, keeping only the even powers of $\bm \delta$. While $P(\bm \delta | \phi)$ is asymmetric due to conversion point uncertainty associated with the Bragg peak, especially for Mom reconstruction, the polarization is intrinsically invariant under 180$^\circ$ rotations. Odd orders of $\bm \delta$ flip sign under these rotations, so they integrate out and we are left with (in index notation)
\begin{equation}
  \begin{split}
    P(\bm r|\phi) = \int d^2 \bm \delta\, P(\bm \delta|\phi) &\bigg[P_\mathrm{mir}(\bm r) + \frac{1}{2!}\delta_i\delta_j \partial_i\partial_jP_\mathrm{mir}(\bm r)\\
    &+\frac{1}{4!}\delta_i\delta_j\delta_k\delta_\ell \partial_i\partial_j\partial_k\partial_\ell P_\mathrm{mir}(\bm r)\bigg]\\
  \end{split}
  \label{eqn:blur-psf-hess}
\end{equation}
where $\delta_i$ are the coordinates of $\bm \delta$ and $\partial_i$ is a derivative with respect to the $i$th coordinate. This integral can be evaluated with a smooth PSF function $P_\mathrm{mir} (\bm r)$, but we use a discrete PSF image so that the derivatives of Eq.~\ref{eqn:blur-psf-hess} must be substituted with convolution by a matrix. For example, the Laplacian of the PSF is $\nabla^2 P_\mathrm{mir}(\bm r) = \int d^2\bm \delta\, P_\mathrm{mir}(\bm r+\bm \delta) \mathsf{K}(\bm \delta)$ where $\mathsf{K}$ is the matrix
\begin{equation}
  \mathsf{K} = \parens{\begin{matrix}
    0&1&0\\1&-4&1\\0&1&0
  \end{matrix}}.
  \label{eqn:laplacian-kernel}
\end{equation}
In our notation, $\mathsf{K}(\bm \delta)$ refers to the matrix entry of $\mathsf{K}$ corresponding to the pixel containing $\bm \delta$. Similarly, we define eight auxiliary $9\times 9$ matrices $\mathsf{Z}_\sigma$, $\mathsf{Q}_\sigma$, $\mathsf{U}_\sigma$, $\mathsf{Z}_k$, $\mathsf{Q}_k$, $\mathsf{U}_k$, $\mathsf{X}_k$, and $\mathsf{Y}_k$ in appendix \ref{app:matrices} which aid in computing the PSF derivatives.

Eq. \ref{eqn:blur-psf-hess} depends on the moments of $P(\bm \delta | \phi)$, which we simplify by assuming that reconstruction error in the parallel and perpendicular directions are uncorrelated. The only surviving moments are the second- and fourth order moments $\sigma_\parallel^2$ and $k_\parallel^4$ of error in the direction parallel to the polarization axis, and similarly $\sigma_\perp^2$ and $k_\perp^4$ in the perpendicular direction. We further assume that the error in the perpendicular direction has mean zero and is Gaussian-distributed, implying that $k_\perp^4 = 3\sigma_\perp^4$.

To summarize, our leakage pattern predictions are controlled by only three parameters: $\sigma_\parallel$ and $\sigma_\perp$ which control the scale of reconstruction error parallel and perpendicular to the polarization axis, and the fourth-order moment $k_\parallel$ which models the non-Gaussianity of reconstruction error in the parallel direction. All three of these parameters depend on the photon energy as discussed in section \ref{sec:fit}.

For an unpolarized source ($\Pi=0$), Eq.~\ref{eqn:iqu-def} predicts 
\begin{subequations}
  \begin{align}
    \begin{split}&I_0(\bm r) = P_\mathrm{mir}(\bm r) \\
      &\qquad+\int d^2\bm \delta\, P_\mathrm{mir}(\bm r+\bm \delta)\parens{\sigma_+^2 \mathsf{Z}_\sigma(\bm \delta) + k_+^4 \mathsf{Z}_k(\bm \delta)}
    \end{split}\\
    &Q_0(\bm r) = \int d^2\bm \delta\, P_\mathrm{mir}(\bm r+\bm \delta)\parens{\sigma_-^2 \mathsf{Q}_\sigma(\bm \delta) + k_-^4 \mathsf{Q}_k(\bm \delta)}\\
    &U_0(\bm r) = \int d^2\bm \delta\, P_\mathrm{mir}(\bm r+\bm \delta)\parens{\sigma_-^2 \mathsf{U}_\sigma(\bm \delta) + k_-^4 \mathsf{U}_k(\bm \delta)}
  \end{align}
  \label{eqn:unpol-iqu}%
\end{subequations}
where $\sigma_\pm^2 = \sigma_\parallel^2 \pm \sigma_\perp^2$ and $k_\pm^4 = k_\parallel^4 \pm k_\perp^4$. These are the contribution of each photoelectron to $I$, $Q$, and $U$ maps. The total maps are exactly the average of these functions over all photoelectrons, which is equal to these functions applied with the parameters of the average photoelectron due to the following argument. Eqs.~\ref{eqn:unpol-iqu} are photoelectron-independent except for the energy dependence of $\sigma_\parallel$, $\sigma_\perp$, and $k_\parallel$. Since Eqs.~\ref{eqn:unpol-iqu} are linear in the energy dependent terms, the $I$, $Q$, and $U$ maps obey superposition. Taking $\sigma_\parallel^2$ to be the spectrum-weighted averages of $\sigma^2_\parallel(E)$ and so on for the other parameters therefore produces the ensemble map. We neglect the energy dependence of the mirror PSF, which becomes important only above 4--5 keV and would violate this argument.

Qualitatively, Eqs.~\ref{eqn:unpol-iqu} demonstrate the effect of polarization leakage. There is an isotropic blur on $I$, equivalent, to first order, to a Gaussian blur with width equal to the quadrature mean of $\sigma_\parallel$ and $\sigma_\perp$. For $\sigma_\parallel \neq \sigma_\perp$, there is also a leakage pattern in $Q$ and $U$ that depends on the derivatives of the PSF. These effects occur even if the source is unpolarized.

If the source is polarized, more terms are added:
\begin{subequations}
  \begin{align}
    I(\bm r) &= I_0(\bm r) + \mu \frac{Q_0(\bm r)Q_\mathrm{src} + U_0(\bm r)U_\mathrm{src}}{2}\label{eqn:iqu-i} \\
    Q(\bm r) &= Q_0(\bm r) + \mu I_0(\bm r)Q_\mathrm{src}\nonumber\\
    &+\mu \frac{X_0(\bm r)Q_\mathrm{src}  + Y_0(\bm r)U_\mathrm{src}}{2} \label{eqn:iqu-q}\\
    U(\bm r) &= U_0(\bm r) + \mu I_0(\bm r)U_\mathrm{src}\nonumber\\
    &+\mu \frac{Y_0(\bm r)Q_\mathrm{src}  - X_0(\bm r)U_\mathrm{src}}{2} \label{eqn:iqu-u}
  \end{align}
  \label{eqn:iqu}%
\end{subequations}
where $Q_\mathrm{src} = \Pi\cos(2\phi_0)$ and $U_\mathrm{src} = \Pi\sin(2\phi_0)$ are the Stokes parameters of the point source. The auxiliary variables
\begin{subequations}
  \begin{align}
      X_0(\bm r) &= -\frac{k_-^4}{4}\int d^2\bm \delta\, P_\mathrm{mir}(\bm r+\bm\delta) \mathsf{X}_k(\bm \delta)\\
      Y_0(\bm r) &= -\frac{k_-^4}{4}\int d^2\bm \delta\, P_\mathrm{mir}(\bm r+\bm\delta) \mathsf{Y}_k(\bm \delta)
    \end{align}
    \label{eqn:x-y}%
\end{subequations}
have also been used. To produce maps, the photoelectron weight $\mu$ must be averaged similarly to $\sigma_\parallel$ above. These terms effectively bias the blurring of $I$ according to the source polarization (Eq.~\ref{eqn:iqu-i}) and add a PSF-distributed source $Q$ and $U$ map to the leakage pattern  with additional corrections (Eqs.~\ref{eqn:iqu-q}, \ref{eqn:iqu-u}). For connection with \citet{bucciantini2023polarisation}, Eqs.~\ref{eqn:iqu} are also expressed in terms of Mueller matrices in appendix \ref{app:mueller}.

Notice that the source polarization effects (Eqs.~\ref{eqn:iqu}) are reduced by $\mu < 1$ while source polarization-independent leakage (Eqs.~\ref{eqn:unpol-iqu}) is not. This highlights the importance of leakage correction, especially when source polarizations are low.

\subsection{Generalization to extended sources}
\label{sec:extended-deconvolution}

When applied to an extended source, the observed leakage pattern is simply a convolution of the point source leakage patterns with the extended source $I_\mathrm{src}(\bm r)$, $Q_\mathrm{src}(\bm r)$, and $U_\mathrm{src}(\bm r)$. More specifically, the extended leakage maps are
\begin{subequations}
  \begin{align}
    I_\mathrm{ext}(\bm r) &= \int d^2\bm x \, I_\mathrm{src}(\bm x) I(\bm r - \bm x)\\
    Q_\mathrm{ext}(\bm r) &= \int d^2\bm x \, I_\mathrm{src}(\bm x) Q(\bm r - \bm x)\\
    U_\mathrm{ext}(\bm r) &= \int d^2\bm x \, I_\mathrm{src}(\bm x) U(\bm r - \bm x)
  \end{align}
\end{subequations}
Leakage effects can be less visually prominent in these extended source observations because the leakage patterns tend to destructively interfere, but must still be modeled and subtracted when mapping an extended source's polarization. In this section, we describe a method to remove the leakage effect and fit for a source polarization.

Usually $I_\mathrm{src}(\bm r)$ of the source is already known---for example, from a \textit{Chandra} observation. The normalized source polarization $q_\mathrm{src}=Q_\mathrm{src}/I_\mathrm{src}$ and $u_\mathrm{src} = U_\mathrm{src}/I_\mathrm{src}$ are unknown quantities, constrained by the observed $q_\mathrm{obs}$ and $u_\mathrm{obs}$ maps. The problem of deriving a source polarization from these maps is an optimization problem with parameters $q_\mathrm{src}$ and $u_\mathrm{src}$, where the function to be minimized is
\begin{equation}
  F = \sum_{\mathrm{detectors}}\int d^2\bm x\, [\Delta q_i(\bm x)^2 + \Delta u_i(\bm x)^2].
\end{equation}
Here, $\Delta q = q_\mathrm{pred} - q_\mathrm{obs}$, and $q_\mathrm{pred}$ and $u_\mathrm{pred}$ include the leakage predictions given $q_\mathrm{src}$ and $u_\mathrm{src}$ estimates. $F$ is therefore a non-negative number that decreases to zero for a perfect fit. If desired, an additional regulatory term can be added to $F$, proportional to
\begin{equation}
  F_\mathrm{reg} = \int d^2\bm x\, [(\nabla^2 q_\mathrm{src})^2 + (\nabla^2 u_\mathrm{src})^2]
  \label{eqn:regulatory}
\end{equation}
to minimize local fluctuations caused by over-fitting to noise.

We spatially bin $q_\mathrm{obs}$ and $u_\mathrm{obs}$ and minimize $F$ using gradient descent. The gradients can be analytically calculated to be
\begin{multline}
  \frac{\partial F}{\partial q_\mathrm{src}(\bm r)} = \sum_\mathrm{detectors} \mu \int d^2\bm x\,
  \frac{I_\mathrm{src}(\bm r)}{I_\mathrm{pred}(\bm x)}\bigg[2\Delta q(\bm x)I_0(\bm x-\bm r)\\
   - Q_0(\bm x - \bm r)\parens{\Delta q(\bm x)q_\mathrm{pred}(\bm x)+\Delta u(\bm x)u_\mathrm{pred}(\bm x)}\\
   +\Delta q(\bm x) X_0(\bm x - \bm r) + \Delta u (\bm x) Y_0 (\bm x - \bm r)\bigg]
  \label{eqn:grad-q}
\end{multline}
and
\begin{multline}
  \frac{\partial F}{\partial u_\mathrm{src}(\bm r)} = \sum_\mathrm{detectors}\mu \int d^2\bm x\,
  \frac{I_\mathrm{src}(\bm r)}{I_\mathrm{pred}(\bm x)}\bigg[2\Delta u(\bm x)I_0(\bm x-\bm r)\\
   - U_0(\bm x - \bm r)\parens{\Delta q(\bm x)q_\mathrm{pred}(\bm x)+\Delta u(\bm x)u_\mathrm{pred}(\bm x)}\\
   +\Delta q(\bm x) Y_0(\bm x - \bm r) - \Delta u (\bm x) X_0 (\bm x - \bm r)\bigg]
  \label{eqn:grad-u}
\end{multline}
The gradient of the regulatory term (Eq.~\ref{eqn:regulatory}) is
\begin{equation}
  \frac{\partial F_\mathrm{reg}}{\partial q_\mathrm{src}(\bm r)} = 2\int d^2\bm x\, (\nabla^2 q_\mathrm{src} (\bm x))\mathsf{K}(\bm x - \bm r)
  \label{eqn:grad-reg}
\end{equation}
and likewise for $u_\mathrm{src}$, where $\mathsf{K}$ is given in Eq.~\ref{eqn:laplacian-kernel}.

Gradient descent proposes that we adjust $q_\mathrm{src}$ and $u_\mathrm{src}$ iteratively, using
\begin{equation}
  q_\mathrm{src}(\bm r) \rightarrow q_\mathrm{src}(\bm r) - \kappa \frac{\partial F}{\partial q_\mathrm{src}(\bm r)}
\end{equation}
and the same for $u_\mathrm{src}$, until $F$ is stationary. The rate $\kappa$ is a tuneable parameter. We set $\kappa = [\delta\bm x \cdot \delta (\nabla F)] / [\delta (\nabla F) \cdot \delta (\nabla F)]$, where $\bm x$ is a vector containing all the binned $q_\mathrm{src}$ and $u_\mathrm{src}$ values, and $\delta \bm x$ is the change in $\bm x$ between the previous iteration and the current one. Similarly, $\delta (\nabla F)$ is the change in the gradient. This choice of $\kappa$ is the Barzilai-Borwein method \citep{barzilai1988two}. As initial conditions, we set $q_\mathrm{src}$ and $u_\mathrm{src}$ equal to the Stokes coefficients observed in detector 1 (divided by the average weight), which has the best PSF.

This algorithm tends to be unstable when a few pixels are much brighter than the rest. It is helpful to reduce the gradient for these pixels only to ensure convergence.

A demonstration of this method on a synthetic nebula is discussed in section \ref{sec:results-extended}. A similar method has been successfully applied to real observations of the Crab Pulsar Wind Nebula (PWN) \citep{wong2023improved}. 

\subsection{Sky-calibrated PSFs}
\label{sec:psfs}
While the MMA PSFs were measured with high resolution on the ground in the Marshall Space Flight Center Stray Light Test Facility \citep[``ground-calibrated PSFs,''][]{2022JATIS...8b6002W}, inevitably the mirror shape will have changed from the different gravity and temperature on orbit, resulting in different PSFs. We extract a new, ``sky-calibrated'' PSF for each detector from select observations of bright, weakly polarized point sources (4U 1820--303, LMC X-1, GX 9+9, GX 301--2) which better match data.

\begin{figure}\
  \includegraphics[width=\linewidth]{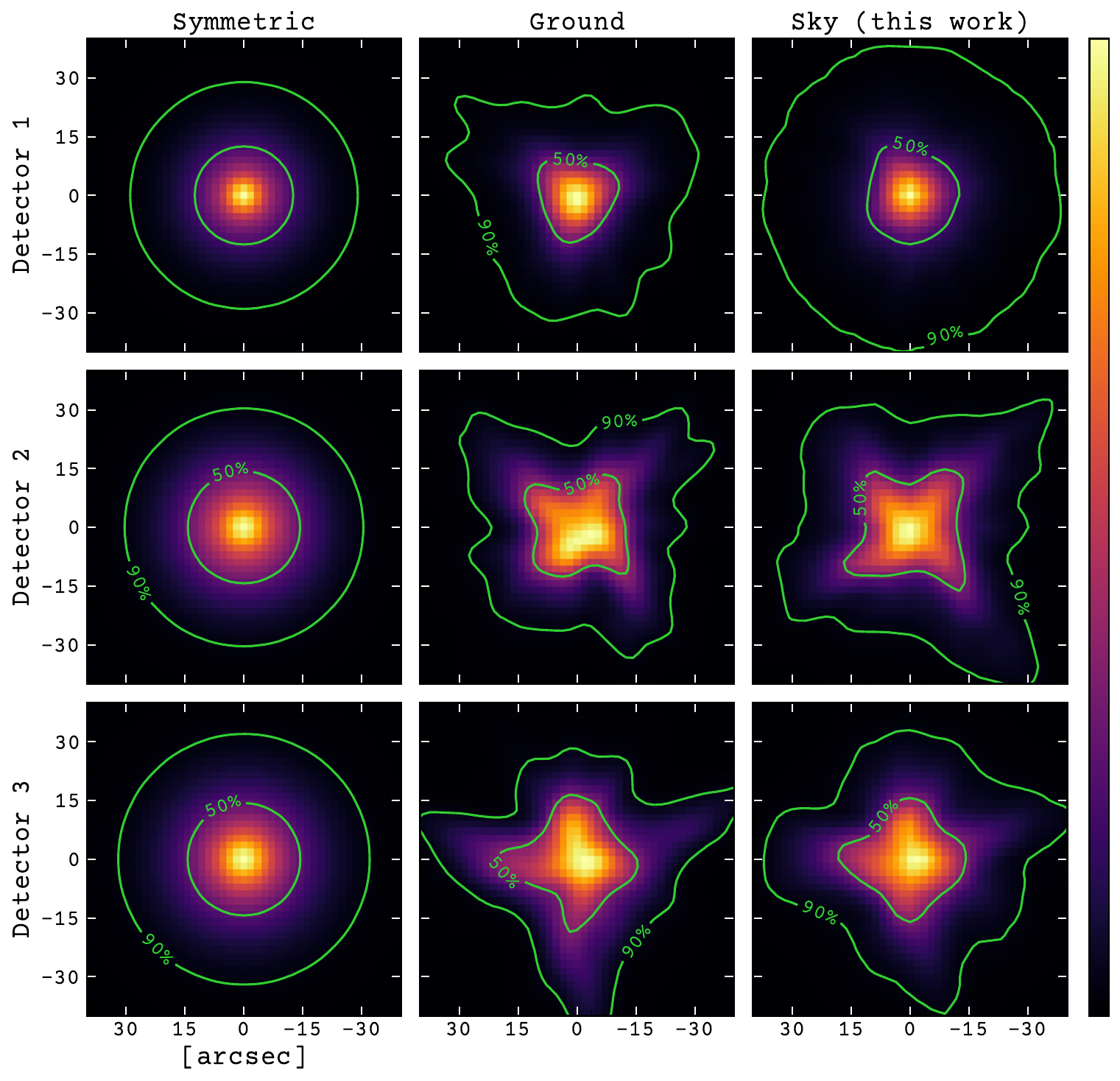}
  \caption{From left to right: the blurred symmetric PSFs used for detector simulations in IXPEobssim, the blurred ground-calibrated PSFs, and our sky-calibrated PSFs. The contours enclose 50\% and 90\% of the flux in the image.}
  \label{fig:psfs}
\end{figure}

We process the level 1 GPD images with the NN-based pipeline to produce $I$ maps for the four sources and fit sky-calibrated PSFs to these maps to remove leakage blur. We set the amplitude of the leakage blur equal to the root mean squared reconstruction error of $10^6$ events simulated by IXPEobssim \citep{Baldini2022obssim} for an unpolarized point source, similarly NN-reconstructed. The resulting PSFs are shown in Fig.~\ref{fig:psfs} with two other commonly used PSFs: the symmetric PSFs used by default in IXPEobssim and the ground-calibrated PSFs.

Our fit method uses gradient descent to minimize the $\chi^2$ value comparing the blurred PSF predicted by Eqs.~\ref{eqn:unpol-iqu} in each detector to observations. We reuse the regulatory term defined in Eq.~\ref{eqn:regulatory} to prevent the amplification of noise. The blurred PSFs---and their $\chi^2$ statistic when matched to data---are analytical functions of the PSF images and can be differentiated by hand as in the previous section.

This PSF fit process can be performed on either Mom- or NN-processed observations. In practice the reduced reconstruction errors in the NN-processed data lead to PSFs that better match \ixpe\ observations, so we use these here.

Though the regulatory term helps to reduce the effect of noise, it also artificially blurs the PSFs. The best accuracy is achieved by using coarse spatial bins so that the Poisson noise in each is low. We use $3^{\prime\prime}$ bins---larger than the pixel scale of the ground PSF map and comparable to the typical error in event position reconstruction. 

The faint wings of the sky-calibrated PSFs are dominated by Poisson fluctuations and PSF extraction becomes inaccurate there. To extend our models beyond $\sim 40''$, we note that the ground PSFs are nearly radial at large angle, dominated by diffraction/scattering spikes. We therefore measure the azimuthally averaged radial profile from each MMA in the ground PSF data and extend the corresponding deconvolved sky PSF core by stretching this radial profile to match the surface brightness of the sky PSF at $\sim 35-40^{\prime\prime}$ in a given azimuthal sector. The profile is then extended to large angle and, after mild smoothing, provides PSF tails matched to our core PSF maps, with appropriate radial and azimuthal intensity. These model PSFs are available in the Zenodo repository linked in section \ref{sec:results}.

\subsection{Fit for the leakage model parameters}
\label{sec:fit}
\begin{figure}
  \centering
  \includegraphics[width=\linewidth]{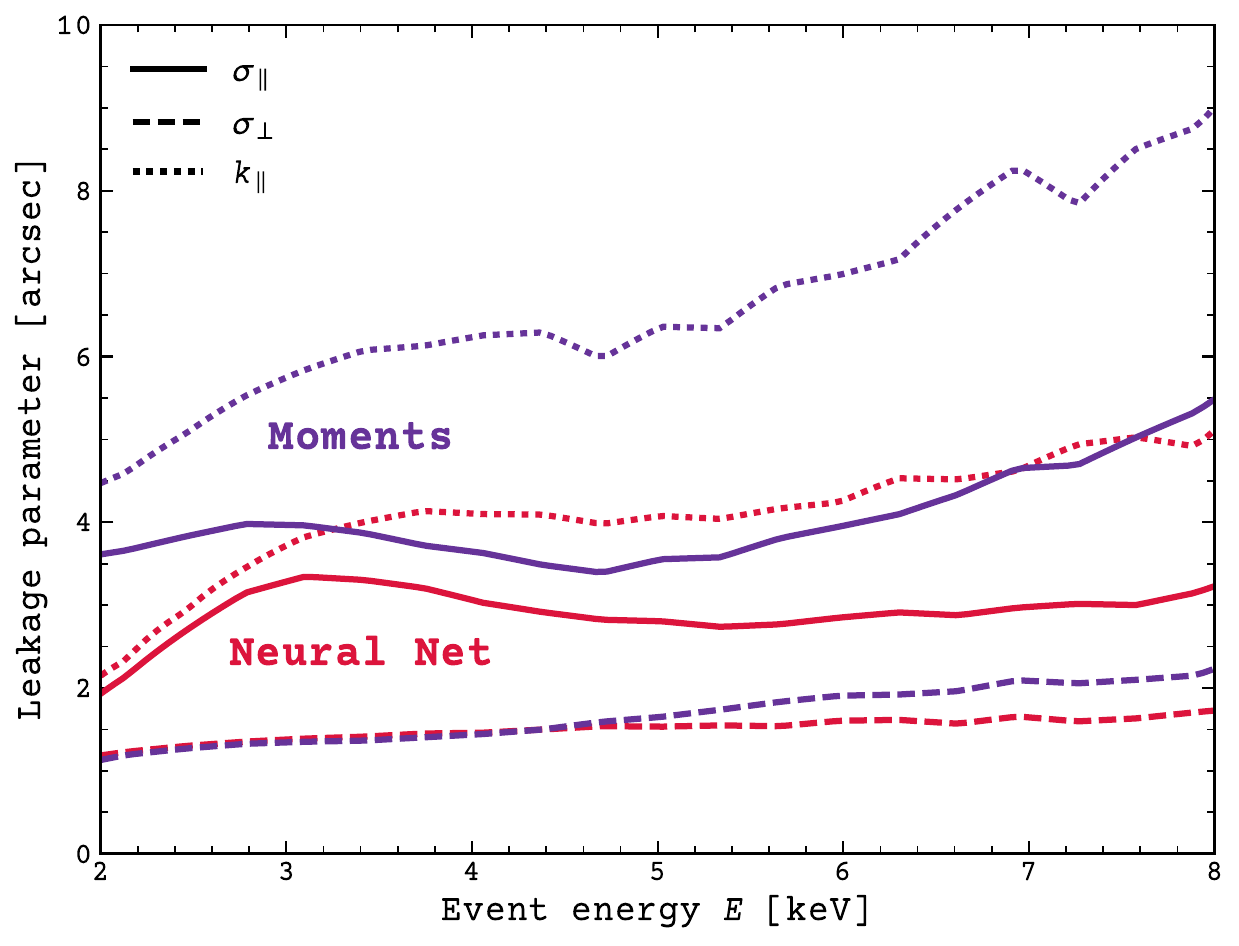}
  \caption{Leakage parameters as a function of event energy $E$ for NN- and Mom-reconstructed data. Energy trends were measured from simulations and slightly rescaled to match observations.}
  \label{fig:energy-dependence}
\end{figure}

The free parameters $\sigma_\parallel$, $\sigma_\perp$, and $k_\parallel$ depend on energy in a way which is difficult to constrain with \ixpe\ data, so we turn to simulated tracks. From these we measure $\sigma_\parallel$, $\sigma_\perp$, and $k_\parallel$ in energy bins roughly 320 eV wide. These data also confirm that perpendicular errors are quite close to Gaussian so that $k_\perp=3^{1/4}\sigma_\perp$ is a good assumption. The simulation does not include effects such as aspecting error which may distort observations and render our leakage maps inaccurate. To reduce this effect, we rescale to $a \cdot \sigma_\parallel(E)$ and $b \cdot k_\parallel(E)$, fitting to observations of our four bright sources with very low phase- and energy-averaged polarization. After spatially binning these data we compute the $\chi^2$ deviation between the observed and predicted $I$, $Q$, and $U$ maps in all pixels within $30''$ of the point sources. Results give $a=1.13\ (1.14)$ and $b=1.06\ (1.00)$ for NN (Mom). These rescaled parameters are shown in Fig.~\ref{fig:energy-dependence}. Rescaling the parameters in this way improves the fit, especially for Mom $Q$ and $U$, though the sky-calibrated PSF outperforms the other two PSFs even without rescaling.

\section{Results \& Discussion}
\label{sec:results}

We have made \texttt{LeakageLib}, our \texttt{Python} software used to predict and remove polarization leakage (sections \ref{sec:leakage} and \ref{sec:extended-deconvolution}), publicly available on Zenodo\footnote{GitHub: \url{https://github.com/jtdinsmore/leakagelib}. Zenodo: \url{https://zenodo.org/records/10483298}}. Leakage prediction for an extended source takes approximately 1 ms per 1,000 pixels on one core, plus overhead for loading and binning the data. Extracting source polarization for extended sources consumes a few minutes per core. The new sky-calibrated PSFs are also available on Zenodo and in the supplemental data of this Paper.

\subsection{Leakage prediction \& data comparison}

\label{sec:results-leakage}
\begin{figure}
  \centering
  \includegraphics[width=0.9\linewidth]{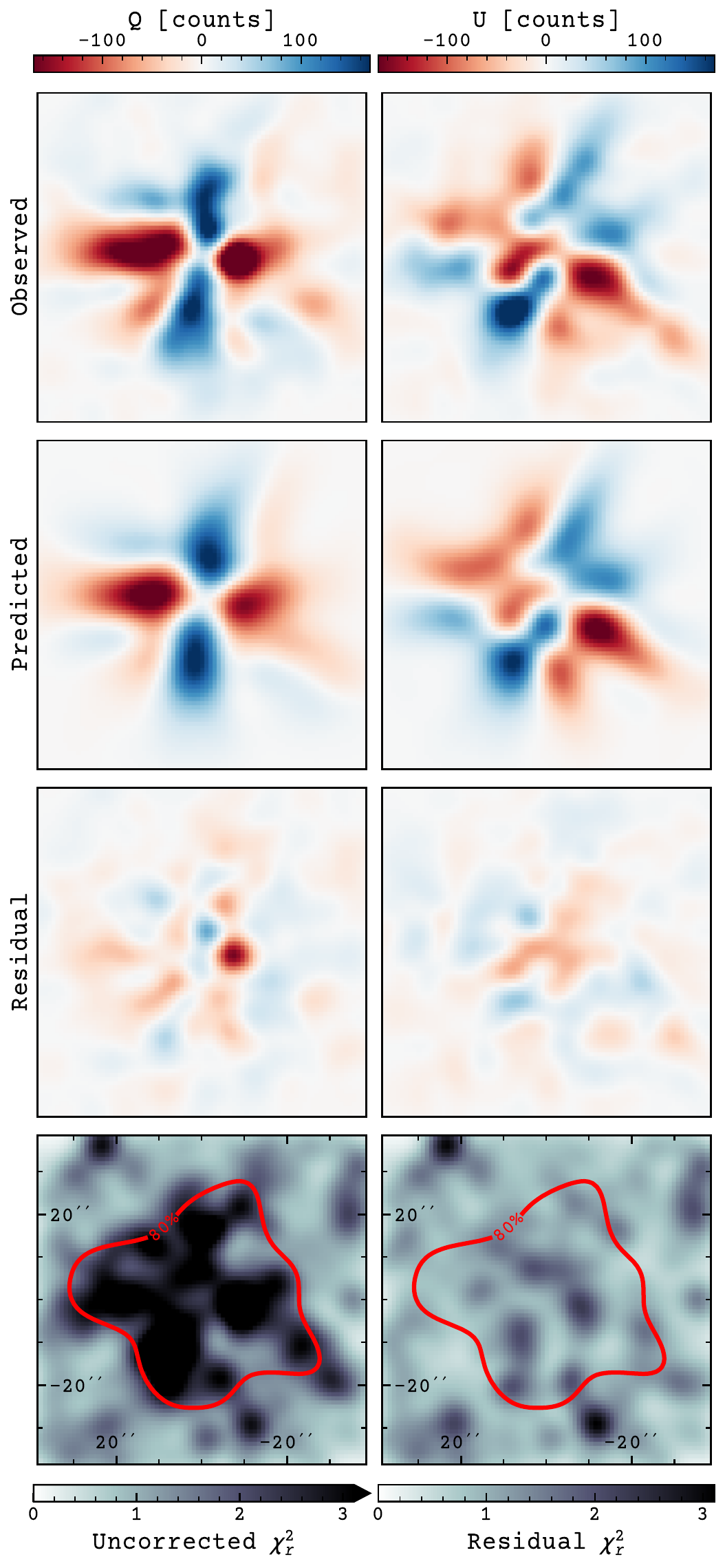}
  \caption{\textit {Row 1:} Polarization leakage detected from the point source 4U 1820$-$303 in DU3. \textit{Row 2:} Leakage predicted by our model. \textit{Row 3:} Residual leakage (data minus prediction). \textit{Row 4:} Significance of the uncorrected observations (\textit{left}) and the residuals (\textit{right}) with an 80\% flux contour from the PSF superimposed. All images are blurred to reduce the visual impact of Poisson noise. Residual leakage is accurately subtracted, even far from the point source.}
  \label{fig:leakage}
\end{figure}

We predict leakage maps for $I$, $Q$, and $U$ for several point sources with weak energy- and time-averaged polarization using our sky-calibrated PSFs and compared the results to observed data. For example, Fig.~\ref{fig:leakage} shows NN-reconstructed 4U 1820$-$303 data as observed by \ixpe\ detector 3. These and all data images shown in this work are unweighted because the correlation between weights, energy, reconstruction error, and polarization introduce small biases to weighted images which we have not yet modeled. The last row shows the significance of the deviation of these polarization maps from zero, obtained using the covariance matrix for $q$ and $u$ \citep{kislat2015analyzing}. The bottom-left panel applies to the uncorrected observed leakage pattern and the bottom-right applies to the residual after leakage correction.

We see that the strong leakage patterns are well predicted by the model and can be subtracted out, leaving minimal residual polarization. The significance of the remaining residuals is near 1, which is the theoretical minimum. Accuracy of leakage prediction is maintained out to the edges of the PSFs, which will be valuable when subtracting extended source polarization underlying the polarization leakage of a bright point source.

\subsection{Improvement with sky-calibrated PSFs}
\label{sec:chisq}
Our leakage model can use any PSF. We compare our sky-calibrated PSFs with alternative symmetric and ground-calibrated PSFs by computing the $\chi^2$ per pixel difference between observed and predicted images, both for NN (top half) and Mom (bottom half) in Table \ref{tab:chisqs}. The upper part of each section lists fit qualities for $I$ images while the lower half characterizes the $Q$ and $U$ residuals. Only pixels within the 80\% flux contour of the image are used, where signal to noise is high.

\label{sec:results-psfs}
\begin{table}
  \centering
  \begin{tabular}{cl|c|cccc}\hline\hline
    &Source & $N\ [10^6]$ & Sym. & Gnd. & Sky & \textbf{Sky*}\\ \hline
    \multirow{4}{0.4em}{\rot{\footnotesize $I$ (NN)~}}
    & 4U 1820 & 1.3& 260 & 330 & 3.1 & \textbf{1.9} \\
    & GX 9+9 & 0.9& 190 & 200 & 2.9 & \textbf{3.5} \\
    & LMC X-1 & 0.9& 160 & 200 & 3.5 & \textbf{2.5} \\
    & GX 301 & 0.2& 23 & 30 & 2.4 & \textbf{2.0} \\
    \hline
    \multirow{4}{0.4em}{\rot{\footnotesize $Q, U$ (NN)~}}
    & 4U 1820 & 1.3& 2.3 & 2.2 & 1.5 & \textbf{1.2}\\
    & GX 9+9 & 0.9& 2.2 & 2.1 & 1.6 & \textbf{1.2}\\
    & LMC X-1 & 0.9& 1.5 & 1.5 & 1.1 & \textbf{1.1}\\
    & GX 301 & 0.2& 1.3 & 1.5 & 1.2 & \textbf{1.1}\\
    \hline\hline
    \multirow{5}{0.4em}{\rot{\footnotesize $I$ (Mom)~}}
    & XTE J1701 & 1.1& 200 & 220 & 5.9 & \textbf{4.9} \\
    & 4U 1820 & 1.1& 220 & 240 & 5.4 & \textbf{4.8} \\
    & GX 9+9 & 0.9& 170 & 180 & 4.4 & \textbf{4.4} \\
    & LMC X-1 & 0.8& 130 & 170 & 7.3 & \textbf{3.4} \\
    & GX 301 & 0.2& 27 & 53 & 3.2 & \textbf{2.0} \\
    \hline
    \multirow{5}{0.4em}{\rot{\footnotesize $Q, U$ (Mom)~}}
    & XTE J1701 & 1.1& 12 & 18 & 3.2 & \textbf{1.5}\\
    & 4U 1820 & 1.1& 13 & 18 & 3.7 & \textbf{1.7}\\
    & GX 9+9 & 0.9& 10 & 14 & 3.0 & \textbf{1.6}\\
    & LMC X-1 & 0.8& 7.2 & 10 & 2.1 & \textbf{1.4}\\
    & GX 301 & 0.2& 5.9 & 11 & 1.8 & \textbf{1.4}\\
  \hline\hline
  \end{tabular}
  \caption{Comparison of predicted point source $I$ and $Q,\,U$ images with \ixpe\ DU123 observations. $N$ gives the average event count per detector, followed by average $\chi^2$/pixel errors (2 significant figures) for the three model PSFs. NN-reconstructed data is shown above the double line with Mom-reconstructed data below. The latter were not used to construct the PSF. In the last column, the leakage parameters have been rescaled with a fit to data. Note that the sky PSF gives significantly improved $Q,U$ models in high count data and dramatically improved $I$ models at all flux levels.}
  \label{tab:chisqs}
\end{table}

If Gaussian blurs with a free scale are added to the symmetric and ground-calibrated PSFs, typical $I$ $\chi^2$ decrease by only about 10. With our sky-calibrated PSFs we obtain a dramatically improved match to $I$, even for Mom, with larger improvements for NN. This is because point source observations are sensitive to PSF asymmetries which are missing in simple symmetric PSFs and are misplaced in the ground-calibrated PSFs, as they are shifted and distorted in the sky-calibrated PSFs (see Fig.~\ref{fig:psfs}). Rescaling the simulated energy dependence yields further improvement.

The $\chi^2/$pixel for the $Q$ and $U$ maps are much lower than for $I$ in all three PSFs due to large $Q$ and $U$ statistical uncertainties, but the sky-calibrated PSFs still out-perform the other PSFs in leakage prediction, especially for the brightest sources. Even small differences such as $\Delta \chi^2_r \approx 0.4$ are significant because these images contain $\sim150$ spatial bins and therefore many degrees of freedom.

For the Mom-reconstructed data we include an additional source, also low-polarization and bright, to demonstrate that our PSFs are not over-fit to the four NN-reconstructed data sets used in their generation.

The linearized polarization leakage treatment of \cite{bucciantini2023polarisation} is similar to our method with the symmetric PSF, but with energy-independent leakage parameters $\sigma_\perp=0$, $\sigma_\parallel\approx 3''$, and $k_\parallel = k_\perp = 0$. With these parameters, the model difference lies in the azimuthal derivative: numerical here and analytic in their case. With $\sigma_\parallel = 3''$ and Mom-reconstructed data, the $\chi^2$ values of the Sym. column of table \ref{tab:chisqs} decrease by $\sim 10$\% for $I$ and $\sim 50$\% for $Q$ and $U$. Our Sky PSF results still provide a large improvement over this previous linear treatment.

The asymmetry of the sky-calibrated PSFs is relevant in measuring the polarization of point sources with small apertures. For Mom-reconstructed data, if the point source polarization is measured within a circular aperture with diameter $d$, polarization leakage contributes an unphysical polarization degree $\mathrm{PD} = $ average of $\sqrt{q^2 + u^2}/\mu_{100}$ of 0.1\% or more in detector 1 if $d < 35''$. Detectors 2 and 3 contribute PD of $>$0.1\% if $d<50''$. To date, apertures used in \ixpe\ analysis have been larger, so the net leakage effect should be negligible; Fig.~\ref{fig:leakage} however emphasizes that non-circular or poorly centroided apertures can induce substantial leakage polarization.

\subsection{Polarization mapping with leakage/PSF models}
\label{sec:results-extended}

\begin{figure*}
  \centering
  \includegraphics[width=\linewidth]{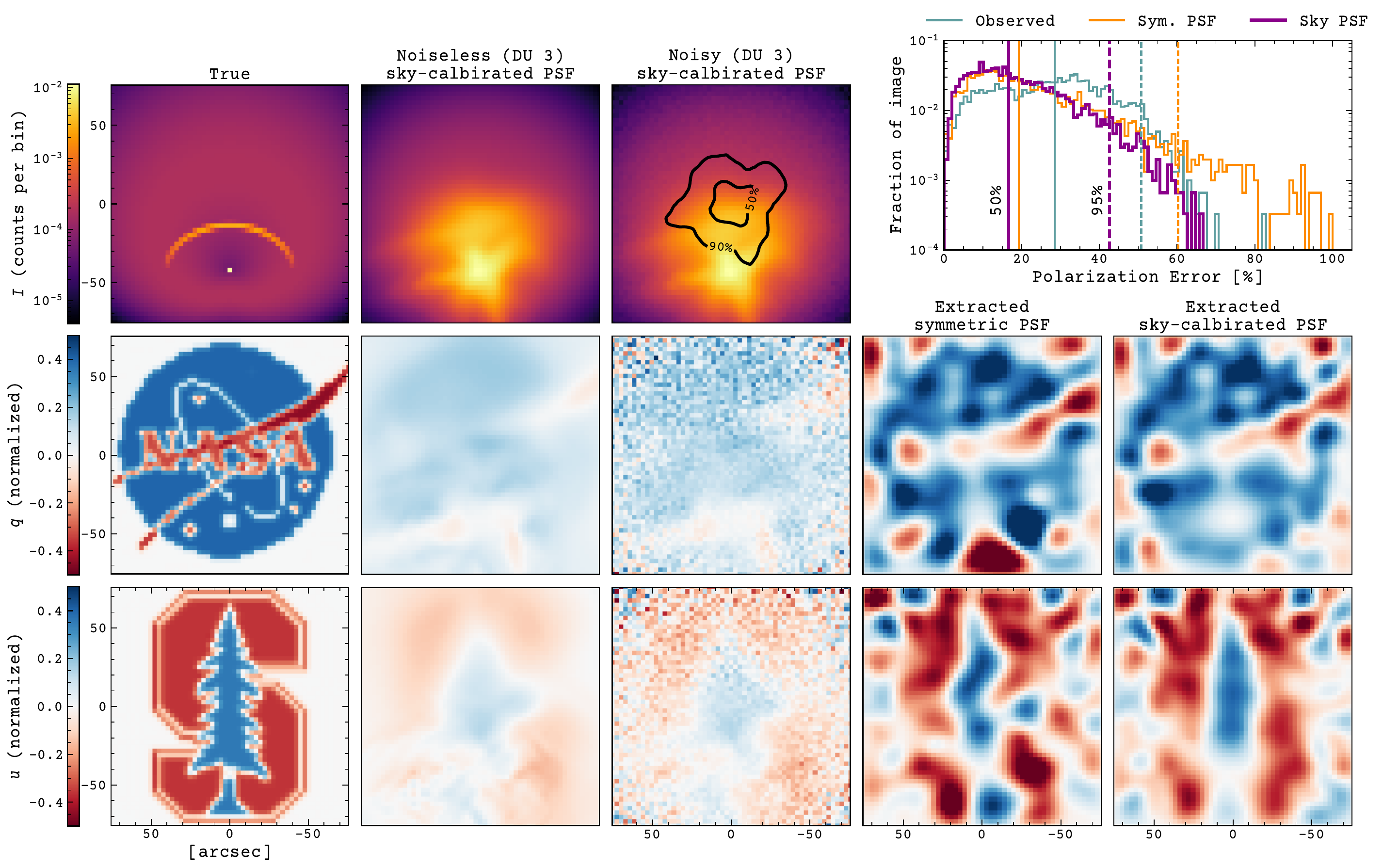}
  \caption{\textit{Left to right}: True flux and polarization of a simulated nebula. Predicted observations using our leakage model for detector 3. The same predictions with Poisson noise added equivalent to a $10^7$-count observation. Extracted source polarizations using the symmetric PSFs. Extracted source polarizations using our sky-calibrated PSFs.  PSF contours are shown for reference in the second column. The sky-calibrated PSFs afford more accurate and more detailed extraction of the source polarization than the inaccurate PSFs, as highlighted in the histogram of polarization errors in the upper right.}
  \label{fig:extended}
\end{figure*}

We test the accuracy of the extended source deconvolution scheme proposed in section \ref{sec:extended-deconvolution} on a synthetic nebula. An unpolarized point source is embedded in a $\sim 40\%$-polarized PWN of equal flux. Our leakage model is then used to blur the model $I$, $Q$, and $U$ maps into simulated \ixpe\ observations. The sky-calibrated PSFs are used, and leakage parameters equivalent to Mom-reconstructed data are assumed. The source is $\sim 2$ arcmin across---about 5 PSF widths. Noise is added equivalent to an observation with 10 million  events per detector, Poisson-distributed in $I$ and distributed according to the $q$-$u$ covariance matrix \citep{kislat2015analyzing} for the polarization maps. This PWN and the simulated observations are shown in the left three columns of Fig.~\ref{fig:extended}.

Our gradient descent method (with regularization) extracts the source $q$ and $u$ images shown in the right two columns of Fig.~\ref{fig:extended} from these noised simulations. We assume that the true $I$ map is available from a well-resolved and high signal-to-noise \textit{Chandra} observation. In one extraction we use the symmetric PSFs; in the other we employ the same sky-calibrated PSFs used to construct the artificial observations.

The method successfully recovers small polarization structures within the nebula. The broad strokes of the NASA text, the red ribbon, the base of the Stanford tree, and the S are all recovered despite not appearing in the raw observations. All of these structures have widths below that of any of the PSFs. However, leakage from the central point source obscures nearby nebula polarization. The maps extracted using the symmetric PSFs (3$^\mathrm{rd}$ column) show the risks of using an inaccurate model. All the major structures of the nebula are recovered, but the leakage of the central point source is incompletely subtracted leaving incorrect nearby polarization predictions. Far from the point source, the image is blotchier and sharp edges in the true polarization map are incorrectly localized.

A more quantitative measure of the extraction's success is presented in the histogram in the upper right of Fig.~\ref{fig:extended} which shows the distribution of polarization error, defined as $\sqrt{(\Delta q)^2 + (\Delta u)^2}$ where $\Delta q = q_\mathrm{true} - q_\mathrm{extracted}$. Both PSFs decrease the average pixel error of the uncorrected data, with sky-calibration giving larger improvement. Moreover the tail of poorly reconstructed pixels with PD error $>50$\% is substantially reduced using sky-PSFs, while symmetric PSFs leave many such errors---more even than the uncorrected data. The sky-calibrated PSFs are therefore valuable to increase overall accuracy and reduce the number of highly inaccurate pixels.

\begin{figure}
  \centering
  \includegraphics[width=\linewidth]{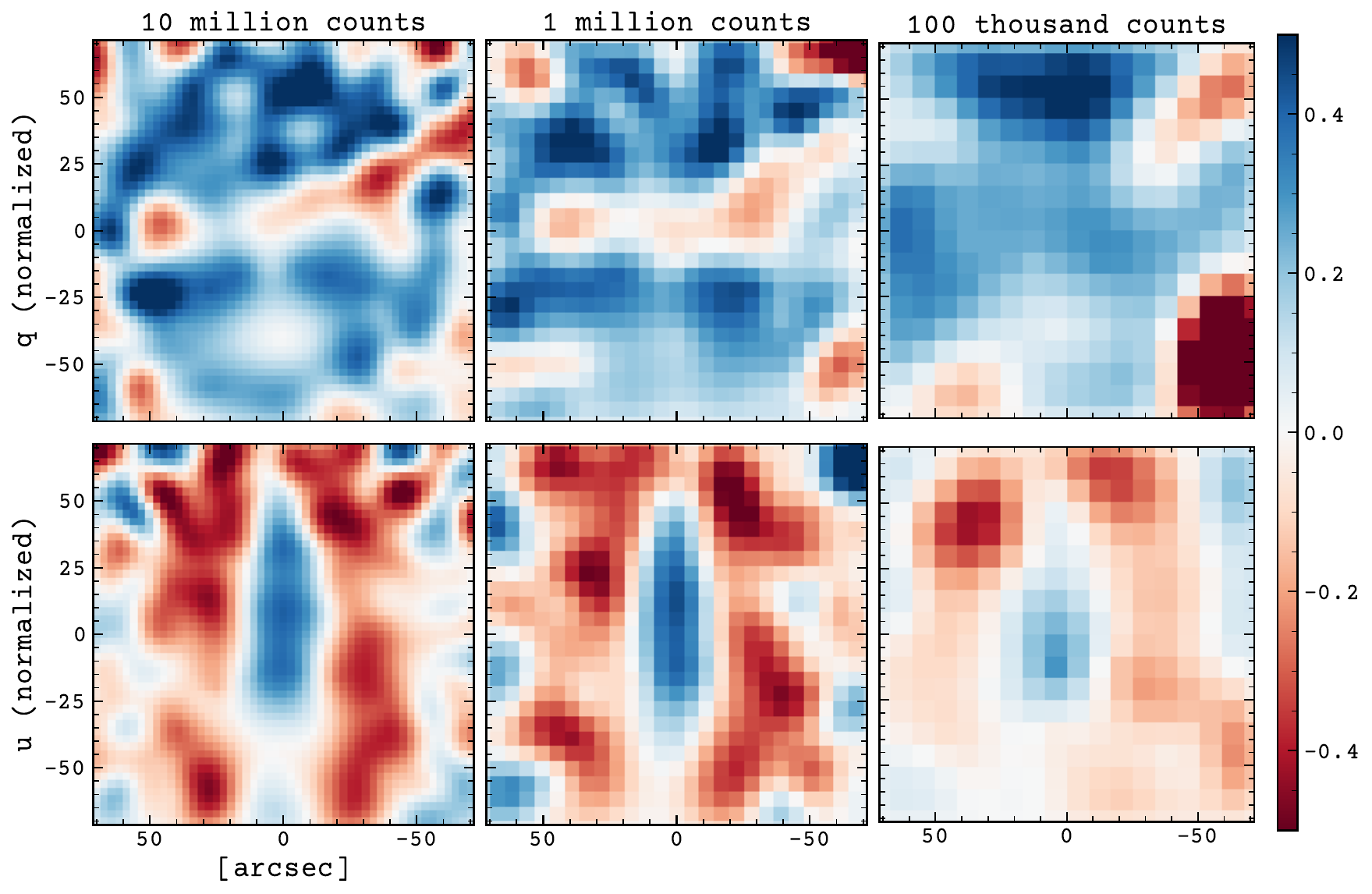}
  \caption{Stokes $q$ and $u$ maps extracted from simulated observations of the synthetic nebula shown in Fig.~\ref{fig:extended}. The observations were noised according to the number of events listed at the top.}
  \label{fig:extended-count}
\end{figure}

Our extraction method's success is aided by the high signal-to-noise ratio afforded by a 10 million count observation. Fig.~\ref{fig:extended-count} compares the source maps extracted from observations with $10^6$ and $10^5$ counts. Over-fitting to the increased noise can be combated by increasing either the weight of the regularization term or the size of the spatial bins, applied as needed. With these shallower observations, the finer details of the source polarization structure are no longer resolved, nor are sharp edges. For the $10^5$ observation, the necessary large regularization term reduces the total polarization degree. Larger scale structures are still detected.

\section{Conclusion}
\label{sec:conclusion}
We have produced sky-calibrated PSFs that match \ixpe\ observations better than alternatives and introduced a polarization leakage model capable of dealing with the PSF asymmetries. Polarization predictions now closely match the leakage patterns observed from data, and the remaining differences are likely due to source-to-source variations in aspecting correction and background levels. As a caveat we note that we have absorbed residual aspecting errors into our fit values for $\sigma_{\parallel}$, $\sigma_\perp$, and $k_\parallel$. As these parameters were derived from bright point sources, the uncorrected aspect errors are small. For fainter sources there will certainly be additional blur from imperfect aspect. Thus while contemporaneous high resolution X-ray images (e.g.\,from \textit{Chandra}) can be used to predict the spatial structure, the models will imperfectly match \ixpe\ fields lacking bright sources for good aspect correction. 

We note that we have so far studied only on-axis source observations and have used 2.3 keV ground calibration PSFs to determine the run of the PSF wings. Further, we have relied on the energy dependence described by IXPEobssim simulated data. This suffices for modeling typical \ixpe\ targets of modest angular extent. However targets with large off-axis flux or with particularly hard spectra could benefit from position- and energy-dependent PSF libraries and checks against high statistics $E>3$\,keV data sets; with these, our leakage prescriptions should still be accurate. We might also include higher order terms in Eq.\,4 to better capture details of the error distribution. A handful of \ixpe\ targets could benefit from a treatment of these effects, but this goes well beyond the present exercise.

Our new PSFs and leakage model can be used to more accurately determine a source polarization map from \ixpe\ observations of extended sources. Small-scale structure and structures near bright point sources, where leakage is most apparent, are particularly improved. Existing \ixpe\ observations of extended sources requiring accurate leakage subtraction include PWNe, supernova remnants, and other complex fields such as the Galactic center and Cen A. Future observations seeking faint polarized structures surrounding bright point sources will benefit from our analysis method. In addition, our improved PSFs can be useful for faint source extraction and extended source modeling.

\section*{Acknowledgments}
  The authors thank Josephine Wong, as well as Niccolo Bucciantini and the rest of the \ixpe\ collaboration for helpful comments. The anonymous referee's comments also spurred upgrade and clarification of the paper's analysis. This work was supported in part by contract NNM17AA26C from the MSFC to Stanford in support of the \ixpe\ project.

\vspace{5mm}
\facilities{IXPE}

\software{HEASoft \citep{nasa2014HEAsoft}, IXPEobssim \citep{Baldini2022obssim}, LeakageLib \citep{dinsmore2024leakagelib}}

\bibliography{ixpepl}{}
\bibliographystyle{aasjournal}

\appendix
\section{Auxiliary Leakage Matrices}
\label{app:matrices}
We defined eight matrices to aid in computing the derivatives of the PSF $P_\mathrm{mir}(\bm r)$ in section \ref{sec:leakage}. They are
\begin{equation}
  \begin{split}
  \mathsf{Z}_\sigma = \frac{1}{(4\Delta)^2}\left(
    \begin{array}{ccccccccc}
     \cdot & \cdot & \cdot & \cdot & \cdot & \cdot & \cdot & \cdot & \cdot \\
     \cdot & \cdot & \cdot & \cdot & \cdot & \cdot & \cdot & \cdot & \cdot \\
     \cdot & \cdot & \cdot & \cdot & 1 & \cdot & \cdot & \cdot & \cdot \\
     \cdot & \cdot & \cdot & \cdot & \cdot & \cdot & \cdot & \cdot & \cdot \\
     \cdot & \cdot & 1 & \cdot & -4 & \cdot & 1 & \cdot & \cdot \\
     \cdot & \cdot & \cdot & \cdot & \cdot & \cdot & \cdot & \cdot & \cdot \\
     \cdot & \cdot & \cdot & \cdot & 1 & \cdot & \cdot & \cdot & \cdot \\
     \cdot & \cdot & \cdot & \cdot & \cdot & \cdot & \cdot & \cdot & \cdot \\
     \cdot & \cdot & \cdot & \cdot & \cdot & \cdot & \cdot & \cdot & \cdot \\
    \end{array}\right)
  \qquad&
  \mathsf{Z}_k = \frac{1}{4(4\Delta)^4}\left(
    \begin{array}{ccccccccc}
     \cdot & \cdot & \cdot & \cdot & 1 & \cdot & \cdot & \cdot & \cdot \\
     \cdot & \cdot & \cdot & \cdot & \cdot & \cdot & \cdot & \cdot & \cdot \\
     \cdot & \cdot & 2 & \cdot & -8 & \cdot & 2 & \cdot & \cdot \\
     \cdot & \cdot & \cdot & \cdot & \cdot & \cdot & \cdot & \cdot & \cdot \\
     1 & \cdot & -8 & \cdot & 20 & \cdot & -8 & \cdot & 1 \\
     \cdot & \cdot & \cdot & \cdot & \cdot & \cdot & \cdot & \cdot & \cdot \\
     \cdot & \cdot & 2 & \cdot & -8 & \cdot & 2 & \cdot & \cdot \\
     \cdot & \cdot & \cdot & \cdot & \cdot & \cdot & \cdot & \cdot & \cdot \\
     \cdot & \cdot & \cdot & \cdot & 1 & \cdot & \cdot & \cdot & \cdot \\
    \end{array}
    \right)\\
  \mathsf{Q}_\sigma = \frac{1}{(4\Delta)^2}\left(
    \begin{array}{ccccccccc}
     \cdot & \cdot & \cdot & \cdot & \cdot & \cdot & \cdot & \cdot & \cdot \\
     \cdot & \cdot & \cdot & \cdot & \cdot & \cdot & \cdot & \cdot & \cdot \\
     \cdot & \cdot & \cdot & \cdot & 1 & \cdot & \cdot & \cdot & \cdot \\
     \cdot & \cdot & \cdot & \cdot & \cdot & \cdot & \cdot & \cdot & \cdot \\
     \cdot & \cdot & -1 & \cdot & \cdot & \cdot & -1 & \cdot & \cdot \\
     \cdot & \cdot & \cdot & \cdot & \cdot & \cdot & \cdot & \cdot & \cdot \\
     \cdot & \cdot & \cdot & \cdot & 1 & \cdot & \cdot & \cdot & \cdot \\
     \cdot & \cdot & \cdot & \cdot & \cdot & \cdot & \cdot & \cdot & \cdot \\
     \cdot & \cdot & \cdot & \cdot & \cdot & \cdot & \cdot & \cdot & \cdot \\
    \end{array}
    \right)
  \qquad&
  \mathsf{Q}_k = \frac{1}{3(4\Delta)^4}\left(
    \begin{array}{ccccccccc}
      \cdot & \cdot & \cdot & \cdot & 1 & \cdot & \cdot & \cdot & \cdot \\
      \cdot & \cdot & \cdot & \cdot & \cdot & \cdot & \cdot & \cdot & \cdot \\
      \cdot & \cdot & \cdot & \cdot & -4 & \cdot & \cdot & \cdot & \cdot \\
      \cdot & \cdot & \cdot & \cdot & \cdot & \cdot & \cdot & \cdot & \cdot \\
      -1 & \cdot & 4 & \cdot & \cdot & \cdot & 4 & \cdot & -1 \\
      \cdot & \cdot & \cdot & \cdot & \cdot & \cdot & \cdot & \cdot & \cdot \\
      \cdot & \cdot & \cdot & \cdot & -4 & \cdot & \cdot & \cdot & \cdot \\
      \cdot & \cdot & \cdot & \cdot & \cdot & \cdot & \cdot & \cdot & \cdot \\
      \cdot & \cdot & \cdot & \cdot & 1 & \cdot & \cdot & \cdot & \cdot \\
    \end{array}
    \right)\\
  \mathsf{U}_\sigma = \frac{1}{(4\Delta)^2}\left(
    \begin{array}{ccccccccc}
     \cdot & \cdot & \cdot & \cdot & \cdot & \cdot & \cdot & \cdot & \cdot \\
     \cdot & \cdot & \cdot & \cdot & \cdot & \cdot & \cdot & \cdot & \cdot \\
     \cdot & \cdot & \cdot & \cdot & \cdot & \cdot & \cdot & \cdot & \cdot \\
     \cdot & \cdot & \cdot & -2 & \cdot & 2 & \cdot & \cdot & \cdot \\
     \cdot & \cdot & \cdot & \cdot & \cdot & \cdot & \cdot & \cdot & \cdot \\
     \cdot & \cdot & \cdot & 2 & \cdot & -2 & \cdot & \cdot & \cdot \\
     \cdot & \cdot & \cdot & \cdot & \cdot & \cdot & \cdot & \cdot & \cdot \\
     \cdot & \cdot & \cdot & \cdot & \cdot & \cdot & \cdot & \cdot & \cdot \\
     \cdot & \cdot & \cdot & \cdot & \cdot & \cdot & \cdot & \cdot & \cdot \\
    \end{array}
    \right)
  \qquad&
  \mathsf{U}_k = \frac{1}{3(4\Delta)^4}\left(
    \begin{array}{ccccccccc}
     \cdot & \cdot & \cdot & \cdot & \cdot & \cdot & \cdot & \cdot & \cdot \\
     \cdot & \cdot & \cdot & -2 & \cdot & 2 & \cdot & \cdot & \cdot \\
     \cdot & \cdot & \cdot & \cdot & \cdot & \cdot & \cdot & \cdot & \cdot \\
     \cdot & -2 & \cdot & 12 & \cdot & -12 & \cdot & 2 & \cdot \\
     \cdot & \cdot & \cdot & \cdot & \cdot & \cdot & \cdot & \cdot & \cdot \\
     \cdot & 2 & \cdot & -12 & \cdot & 12 & \cdot & -2 & \cdot \\
     \cdot & \cdot & \cdot & \cdot & \cdot & \cdot & \cdot & \cdot & \cdot \\
     \cdot & \cdot & \cdot & 2 & \cdot & -2 & \cdot & \cdot & \cdot \\
     \cdot & \cdot & \cdot & \cdot & \cdot & \cdot & \cdot & \cdot & \cdot \\
    \end{array}
    \right)\\
  \mathsf{X}_k = \frac{1}{3(4\Delta)^4}\left(
    \begin{array}{ccccccccc}
     \cdot & \cdot & \cdot & \cdot & -1 & \cdot & \cdot & \cdot & \cdot \\
     \cdot & \cdot & \cdot & \cdot & \cdot & \cdot & \cdot & \cdot & \cdot \\
     \cdot & \cdot & 6 & \cdot & -8 & \cdot & 6 & \cdot & \cdot \\
     \cdot & \cdot & \cdot & \cdot & \cdot & \cdot & \cdot & \cdot & \cdot \\
     -1 & \cdot & -8 & \cdot & 12 & \cdot & -8 & \cdot & -1 \\
     \cdot & \cdot & \cdot & \cdot & \cdot & \cdot & \cdot & \cdot & \cdot \\
     \cdot & \cdot & 6 & \cdot & -8 & \cdot & 6 & \cdot & \cdot \\
     \cdot & \cdot & \cdot & \cdot & \cdot & \cdot & \cdot & \cdot & \cdot \\
     \cdot & \cdot & \cdot & \cdot & -1 & \cdot & \cdot & \cdot & \cdot \\
    \end{array}
    \right)
  \qquad&
  \mathsf{Y}_k = \frac{1}{3(4\Delta)^4}\left(
    \begin{array}{ccccccccc}
     \cdot & \cdot & \cdot & \cdot & \cdot & \cdot & \cdot & \cdot & \cdot \\
     \cdot & \cdot & \cdot & 4 & \cdot & -4 & \cdot & \cdot & \cdot \\
     \cdot & \cdot & \cdot & \cdot & \cdot & \cdot & \cdot & \cdot & \cdot \\
     \cdot & -4 & \cdot & \cdot & \cdot & \cdot & \cdot & 4 & \cdot \\
     \cdot & \cdot & \cdot & \cdot & \cdot & \cdot & \cdot & \cdot & \cdot \\
     \cdot & 4 & \cdot & \cdot & \cdot & \cdot & \cdot & -4 & \cdot \\
     \cdot & \cdot & \cdot & \cdot & \cdot & \cdot & \cdot & \cdot & \cdot \\
     \cdot & \cdot & \cdot & -4 & \cdot & 4 & \cdot & \cdot & \cdot \\
     \cdot & \cdot & \cdot & \cdot & \cdot & \cdot & \cdot & \cdot & \cdot \\
    \end{array}
    \right)
  \end{split}
\end{equation}
where $\Delta$ is the width of pixels in the PSF image and dots indicate zeros.

\section{Mueller Matrices}
\label{app:mueller}
Mueller matrices \citep{tinbergen1996astronomical} are a mathematical tool to efficiently express how polarization leakage patterns depend on the source polarization. If a point source is located at position $\bm r'$ with Stokes parameters $I_\mathrm{src}$, $Q_\mathrm{src}$, $U_\mathrm{src}$, then the observed polarization at another position is given by
\begin{equation}
  \parens{\begin{matrix}
    I(\bm r) \\Q(\bm r) \\ U(\bm r)
  \end{matrix}} = \int d^2\bm r'\, M(\bm r - \bm r')
  \parens{\begin{matrix}
    I_\mathrm{src}(\bm r') \\Q_\mathrm{src}(\bm r') \\ U_\mathrm{src}(\bm r')
  \end{matrix}}
  \label{eqn:mueller-def}
\end{equation}
for Mueller matrix $M(\bm r - \bm r')$. Eqs.~\ref{eqn:iqu} for an unpolarized point source dictates that
\begin{equation}
  M(\bm x) = \parens{\begin{matrix}
    I_0(\bm x) & \frac{\mu}{2} Q_0(\bm x) & \frac{\mu}{2}U_0(\bm x)\\
    Q_0(\bm x) & \mu\brackets{I_0(\bm x)- \frac{1}{2}X_0(\bm x)} & \frac{\mu}{2} Y_0(\bm x)\\
    U_0(\bm x) & \frac{\mu}{2} Y_0(\bm x) & \mu \brackets{I_0(\bm x) + \frac{1}{2}X_0(\bm x)}\\
  \end{matrix}}
\end{equation}
using functions defined in \ref{eqn:unpol-iqu} and \ref{eqn:x-y}. Note that $M(\bm x)$ can be computed directly from the PSF without any prior knowledge of the source aside from the parameters $\sigma_\parallel$, $\sigma_\perp$, $k_\parallel$, and $\mu$, which depend on the source spectrum. Polarization maps can be computed for an extended source by convolving $M$ with the source polarization maps---i.e., by integrating the left hand side of \ref{eqn:mueller-def} with respect to $\bm r'$.

\end{document}